# Physical and geographic analysis of the urban cooling potential

*Martin Hendel[1],*
*Univ Paris Diderot, Sorbonne Paris Cité, LIED, UMR 8236, CNRS (France)*
*Université Paris-Est, ESIEE Paris, département SEN, (France)*

*Sophie Parison,*
*Paris City Hall, Water and Sanitation & Roads and Traffic Divisions (France)*
*Univ Paris Diderot, Sorbonne Paris Cité, LIED, UMR 8236, CNRS (France)*

*Laurent Royon,*
*Univ Paris Diderot, Sorbonne Paris Cité, LIED, UMR 8236 (France)*

**ABSTRACT**

The performance of a number of urban cooling techniques has been thoroughly studied by the scientific community. However, decision-makers lack the tools to spatialize their deployment as part of their urban cooling and climate change adaptation strategies. Among other indicators, a spatial assessment of the cooling potential for a technique in a given area is lacking. To this end, we analyze the physical mechanisms on which these techniques are based and identify corresponding geographical indicators that influence their cooling performance. Solar irradiance, existing material properties and underground infrastructure stand out as essential indicators for this purpose.

## Nomenclature

| Symbol | Quantity | [SI unit] |
|---|---|---|
| $\alpha$ | albedo | [-] |
| $\epsilon$ | emissivity | [-] |
| $E$ | evaporation rate | [kg.s$^{-1}$] |
| $H$ | sensible (convective) heat flux density | [W.m$^{-2}$] |
| $l$ | latent heat of vaporization of water | [kJ.kg$^{-1}$] |
| $L$ | LW radiation | [W.m$^{-2}$] |
| $R_n$ | net radiation | [W.m$^{-2}$] |
| $\sigma$ | Boltzmann constant | [W. m$^{-2}$.K$^{-4}$] |
| $S$ | SW radiation | [W.m$^{-2}$] |
| $T$ | surface temperature | [°C] |

| Abbreviation | Description |
|---|---|
| GIS | geographic information system |
| LW | longwave (3-100 µm) |
| PCM | phase-change materials |
| SW | shortwave (0,3-3 µm) |
| UHI | urban heat island |
| up | upwards |

---

[1] Corresponding Author: martin.hendel@univ-paris-diderot.fr.

# Introduction

The goal of countering the effects of heat waves and urban heat islands (UHI) has been adopted by many cities around the world. This awareness is accelerated by climate change forecasts, which predict an increase in these phenomena. Depending on the considered greenhouse gas emission scenario, some cities are even facing the risk of becoming uninhabitable all or part of the year (Kang & Eltahir, 2018; Mora et al., 2017; Pal & Eltahir, 2015). The interest of urban decision-makers in cooling techniques is therefore growing and has stimulated research.

Many efforts have been made to study the performance of various cooling techniques. This includes, for example, cool materials, which include reflective, emissive, permeable or PCM materials (Santamouris, 2015; Taha, 1997). Urban greening has also been widely studied with numerous studies focusing on the benefits of planting trees, developing green facades or roofs, planters, and greening tree bases or creating parks (Bowler, Buyung-Ali, Knight, & Pullin, 2010; Taha, 1997). Energy efficiency measures are also important, particularly those that seek to limit the impact of air conditioning (Oke, 1982; Taha, 1997; Tremeac et al., 2012). Finally, less common techniques such as urban watering have also been studied (Hendel, Gutierrez, Colombert, Diab, & Royon, 2016; Kinouchi & Kanda, 1997; Takahashi, Asakura, Koike, Himeno, & Fujita, 2010; Yamagata, Nasu, Yoshizawa, Miyamoto, & Minamiyama, 2008).

These techniques are nowadays generally categorized by their place of application, i.e. roof, façade or pavement. This work is useful in making solutions available to urban decision-makers, however they lack the tools to build an urban cooling strategy over their jurisdiction. Specifically, it is difficult to integrate these solutions into spatial analyses without the need for generally slow, cumbersome and delicate simulations, often outside the technical capabilities of these actors. On the other hand, geographic information system (GIS) tools are very common in urban studies, but there is little work to bridge the gap between known physics and this type of tool.

To this end, this contribution aims to identify readily-available geographical indicators that can be used in GIS tools to assess the cooling potential of a given area for hot weather conditions. First, the urban energy balance is analyzed to discriminate techniques according to the physical phenomena on which they rely rather than on their place of application. From this, the urban geographical parameters best able to transcribe them are identified. These geographical parameters can then be used to calculate an indicator of urban cooling potential for a given area for various techniques using standard GIS tools.

# Methodology

The cooling potential indicator aims to identify areas with favorable conditions suited to obtaining the highest performance from cooling techniques. It is therefore necessary to identify these conditions and the urban parameters that make it possible to evaluate them. Several parameters may be necessary and they are dependent on the cooling technique considered. We begin by analyzing the energy balance of an urban facet and an urban volume to identify the physical phenomena on which cooling techniques are based and classifying them accordingly.

**Urban facet energy balance**

The urban energy balance can be analyzed at the scale of an urban facet, i.e. a roof, facade or pavement, as shown in Figure 1.

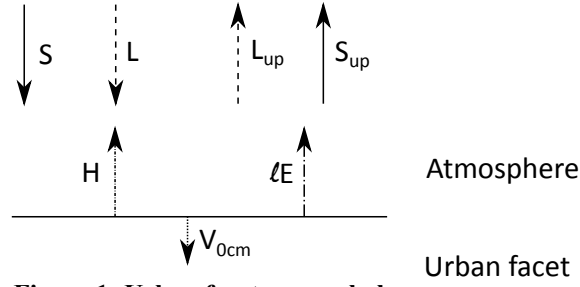

**Figure 1: Urban facet energy balance**

The term $H$ represents the convective exchange between the ground and the atmosphere; $V$ is surface conduction into the ground; $lE$ is the latent heat flow due to the (possible) evaporation of water present at the surface. The latter is the product of the latent heat of water vaporization $l$ and the evaporation rate $E$. Incident radiative fluxes are noted $S$ and $L$ for short and long wavelengths respectively (SW and LW) and $S_{up}$ and $L_{up}$ for SW and LW radiosity, respectively.

We note $R_n$ the net radiation, i.e. the radiative balance of incident radiation and radiosity:

$$R_n = S + L - L_{up} - S_{up} \tag{1}$$

Furthermore, the energy conservation principle gives us:

$$R_n = H + lE + V \tag{2}$$

In other words:

$$S + L = \alpha S + (1 - \epsilon)L + \epsilon \sigma T^4 + H + lE + V \tag{3}$$

In this last equation, the term on the left is incident radiation, i.e. the inbound flows, while the terms on the right represent the outbound flows. The latter depend on the surface's radiative properties (albedo $\alpha$ and emissivity $\epsilon$) and surface temperature for the radiative portion, while the other terms depend on convective and conductive exchange coefficients and temperature or vapor pressure gradients, which itself is dependent on surface temperature. Albedo and emissivity are fixed parameters, resulting from the choice of material or paint, therefore the remaining (non-reflective) radiative, convective and conductive heat flows change during the day to balance out incident radiation.

**Urban volume energy balance**

Figure 2 illustrates the energy balance of an urban volume whose upper limit is above the urban canopy layer and whose depth is such that the conductive heat flux is negligible over the time scale considered. $Q_F$ refers to the release of anthropogenic atmospheric heat; $\Delta Q$ is the heat storage term within the urban materials present in the volume. We note $Q_A$ the term for heat advection outside the volume. This exchange can be horizontal, for example due to wind or river transportation, but can also be vertical, e.g. in the case of geothermal energy.

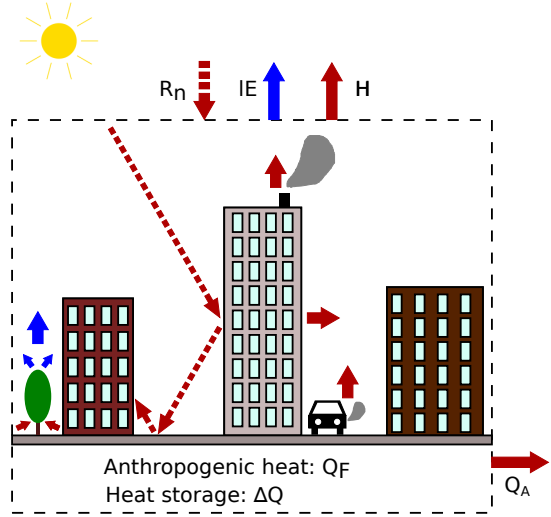

**Figure 2: Urban volume energy balance**

The energy conservation equation for the urban volume is:

$$R_n + Q_F = H + lE + \Delta Q + Q_A \quad (4)$$

By expliciting the contents of $R_n$, a similar form to equation (3) with incident flows on the left and outbound flows on the right can be obtained for an urban volume:

$$S + L + Q_F = \alpha S + (1 - \epsilon)L + \epsilon \sigma T_s^4 + H + lE + \Delta Q + Q_A \quad (5)$$

**Incident radiation partitioning**

The analysis of the partition of net radiation (i.e. absorbed radiation) between convective, latent and conductive flux ($H$, $E$ and $V$) was initiated for an agricultural area by Camuffo and Bernardi (1982) and subsequently adapted to the study of urban materials (Anandakumar, 1999; Asaeda, Ca, & Wake, 1996; Qin & Hiller, 2014).

We adopt this approach and extend it to analyze the partitioning of the inbound flows, i.e. incident radiation ($S+L$) and anthropogenic releases ($Q_F$), for an urban volume. This allows us to make clear the physical role of urban radiative properties (albedo α and emissivity ϵ).

Table 1 and Table 2 illustrate the relative importance of the heat flow and radiative terms found in equation (5) for a mid-latitude city under clear skies in summer.

**Table 1: Clear sky summertime heatflows for a mid-latitude city in W/m²** (Oke, 1988, 1997).

|  | $R_n$ | $Q_F$ | $lE$ | $H$ | $\Delta Q$ |
|---|---|---|---|---|---|
| Day | 516 | 30 | 158 | 240 | 148 |
| Night | -80 | 20 | 13 | 7 | -80 |

Table 2: Clear sky summertime radiative balance for a mid-latitude city in W/m² (Oke, 1988, 1997).

|       | S   | L   | $S_{up}$ | $L_{up}$ |
|-------|-----|-----|----------|----------|
| Day   | 760 | 365 | 106      | 503      |
| Night | 0   | 335 | 0        | 415      |

Clearly, the most important incoming flux is net radiation $R_n$, more specifically solar irradiance $S$.

**Urban cooling mechanisms**

Regardless of the scale considered, it is the sensitive heat exchanges ($H$) that are responsible for the increase in air temperature observed in cites. UHI countermeasures therefore aim to reduce them by acting on the other terms of the energy balance (Christen & Vogt, 2004), either by reducing inbound energy flows or increasing outbound flows other than $H$.

On this basis, the following classification of cooling mechanisms can be proposed according to the terms of the energy balance that are modified:

- **Reduce inbound flows**
    1. Improve the urban radiative balance (reduction of $R_n$)
        a. via shading (reduction of $S$)
        b. via higher reflectivity (increase $S_{up}$)
        c. via higher emissivity (increase $L_{up}$)
        d. via lower LW radiation (decrease $L$)
    2. Decrease in atmospheric anthropogenic heat release
        a. energy efficiency (decrease $Q_F$)
        b. heat sink transfer (from $Q_F$ to $Q_A$)
- **Increase outbound flows**
    3. Increase latent flows ($lE$)
    4. Increase heat storage ($V$ and/or $\Delta Q$)
        a. via heat harvesting
        b. via modified thermal properties (e.g. thermal conductivity and inertia)
        c. other methods

Mechanisms 1 and 2 seek to reduce the incoming flows $R_n$ and $Q_F$. As described above, it does not matter which of the terms of the radiation balance is affected: (a) reduction of solar irradiance $S$, (b) increase of SW radiosity $S_{ref}$ or (c) increase of LW radiosity ($L_{up}$) or (d) decrease of LW incident radiation ($L$). This last lever is not generally mentioned in the literature separately from shading. Indeed, shading devices do not necessarily contribute to reducing $L$, especially when the temperature of the shading device rises due to solar heating. However, there are devices that address both simultaneously (S Sakai et al., 2012; Satoshi Sakai, Sugawara, Misaka, Narita, & Honjo, 2018). Nevertheless, we leave this mechanism aside here.

Mechanism 2 aims to reduce the source term $Q_F$, either by improving energy efficiency or by transferring atmospheric emissions to other sinks, for example via geothermal air conditioning, which results in a decrease of $Q_F$ in exchange for an increase in $Q_A$.

Mechanisms 3 and 4 aim to increase outflows, i.e. provide an alternative outbound heat flow to sensible heat. In essence, they seek to absorb source heat (solar or anthropogenic) to evaporate water (Mechanism 3) or for energy use or storage (Mechanism 4). Mechanism 3 focuses on increasing latent flows. Mechanism 4, on the other hand, includes changes in heat conduction or storage. These may include (a) the harvesting of solar energy, for example absorbed by heat exchangers integrated into a pavement or by thermal or photovoltaic solar collectors. Finally, it includes (b) techniques to increase the thermal inertia of materials, with or without the incorporation of phase change materials. While this does not modify the energy balance, it serves to smooth out the intensity of heat flows by lowering urban facet surface temperatures.

It should be noted that in reality most cooling techniques involve a combination of these mechanisms. This is the case, for example, for trees that provide shade during the day, but also limit LW radiosity at night by changing the sky view factor in addition to increasing latent flows by evapotranspiration.

Table 3 provides a classification of the most common cooling techniques by their principal mechanism.

Table 3: Classification of standard approaches to urban cooling by mechanism

| Mechanism | Techniques |
|---|---|
| 1 | Reflective and/or emissive materials |
| 2 | Energy efficiency, aqua- or geothermal energy |
| 3 | Urban vegetation (additional parks or trees), green materials (facades or roofs), permeable materials |
| 4 | Solar panels, heat exchangers incorporated in coatings, urban materials incorporating MCP or other meta-materials |

# Results

From the previous classification of the physical mechanisms on which cooling techniques are based, we now identify the GIS information or data that may correspond to them.

**Solar irradiance (mechanisms 1a, 1b, 3 and 4)**

As shown in Table 2, this is the most important inbound flow and affects the most cooling mechanisms. Indeed, mechanisms 1a, 1b, 3 and 4 are concerned since they aim to reduce solar gains, i.e. avoid the absorption of solar energy in the form of sensible heat or change its form. This is true of shading devices (1a), cool materials (1b), or evaporative cooling via permeable materials or by greening (3) whose performance is highest under direct sunlight. The same applies to harvesting solar energy or its storage in phase change materials (4). In other words, only techniques which rely on mechanisms 1c or 2 are not directly affected by the intensity of solar irradiance. This makes the solar irradiance of a targeted area an essential parameter to evaluate the cooling potential of cooling techniques based on mechanisms 1a, 1b, 3 and 4.

Clear-sky surface irradiance in a neighborhood or city can be simulated via ray-tracing from a digital elevation model (DEM) of the study area, i.e. a 3D map, for any time of the year, specifically in summer. This type of simulation is already available in some popular GIS tools (e.g. SAGA). In the field, the more recent the DEM is and the higher its spatial resolution is, the more reliable the simulation will be. Figure 3 illustrates the surface irradiance of the Alesia district in

Paris on August 16th between 10am and 2pm (solar time) as simulated from a 2012 DEM with a spatial resolution of 0.5 m.

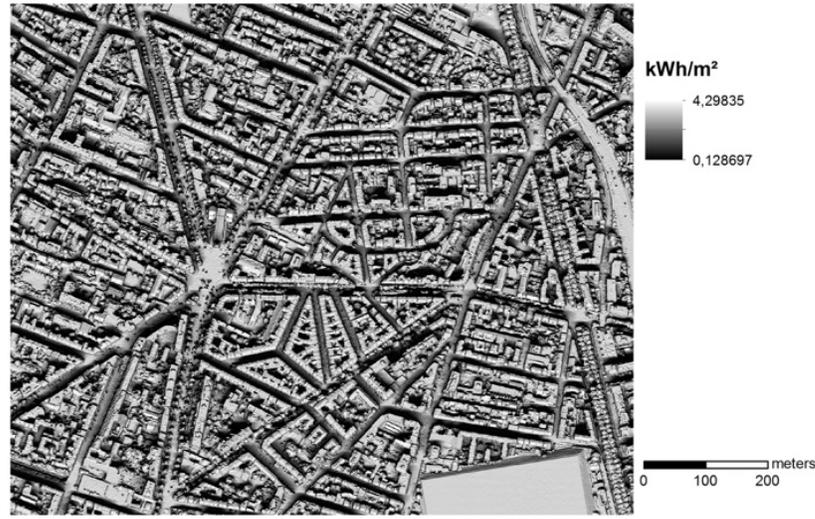

Figure 3: Cumulative solar irradiance between 10 am and 2 pm (local solar time) simulated on the 16th of August

By setting an appropriate threshold, for example 50% of global horizontal irradiance during the considered period, it is thus possible to immediately identify areas with high cooling potential from this type of information.

**Existing material properties (mechanisms 1, 3 and 4)**

The orders of magnitude cited in Table 1 and Table 2 are valid for the average thermal (thermal conductivity and capacity) and radiative (albedo and emissivity) properties of materials encountered in cities. Of course, they depend on the materials and morphology actually present at a finer scale. In other words, the gains to be expected from techniques under mechanisms 1, 3 and 4 depend on the materials already present that will be replaced or treated. Obviously, reducing solar gains from low albedo materials will be more effective than from high albedo materials. This also applies to permeable materials that will more advantageously replace impermeable materials.

This type of information is not widely available for many urban areas, Paris being no exception. Nevertheless, when available, this information is useful in assessing the cooling potential.

**Underground infrastructure (mechanisms 1b, 1c, 3 and 4)**

The presence of underground infrastructure can locally affect the heat balance in a positive or negative way by adding a source term that affects the term $V$. This mainly concerns shallow infrastructure (a few meters deep at most) such as district heating (negative local impact) or cooling (positive local impact) networks or shallow high-traffic metro stations for which the daily average of $V$ is not equal to zero.

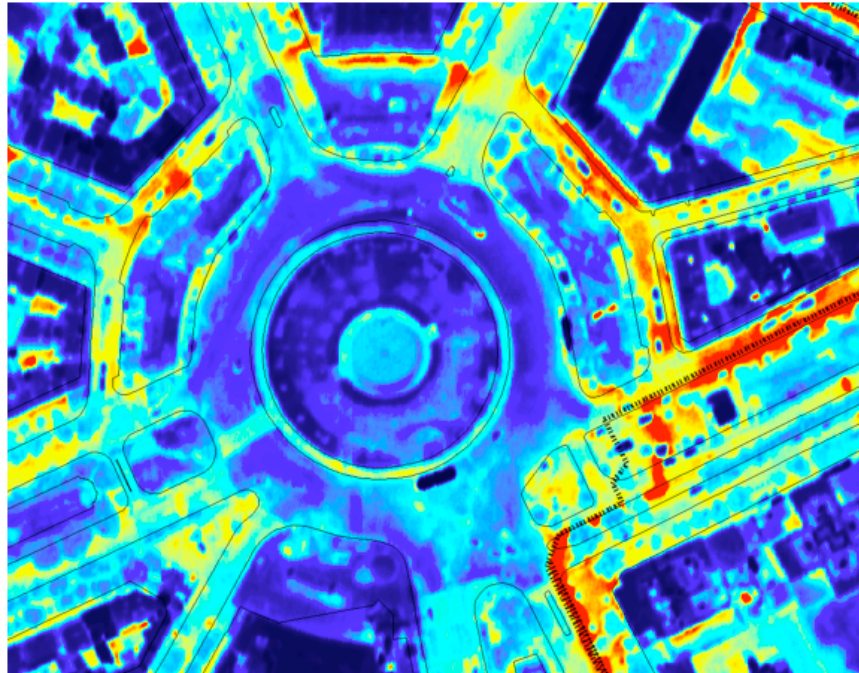
**Figure 4:** Airborne infrared photograph of Place d'Italie in Paris taken on the night of August 18$^{th}$, 2015 under clear sky conditions.

The infrared thermography of Place d'Italie in Paris presented in Figure 4 provides some concrete examples of this. In the East, the impact of the heating network (dotted black line) has a visible impact on surface temperatures. On the northern half of the outer circumference of the Place, it is the thermal impact of metro Line 5 that is visible.

Urban decision-makers generally have maps of heating or cooling networks or metro stations, which facilitates their integration into this type of approach, even if their impact is very localized.

**Other parameters**

Several other parameters can be taken into account though their importance may be specific to certain cooling techniques.
- **Sky view factor (mechanisms 1b and 1c)**

Radiative cooling techniques (mechanisms 1b and 1c) reflect or re-emit the solar energy received back towards the sky, outside the urban volume. Their effectiveness therefore depends on the sky view factor of the treated surfaces, whether they are horizontal (roofs or pavements) or vertical (facades). As a result, it is clear that roofs are the surfaces most likely to benefit from these techniques, benefiting from both high solar irradiance (for 1b only) and high sky view factor.

As with solar irradiance, this parameter can be calculated on the basis of a DEM.
- **Relative humidity (mechanism 3)**

Techniques based on water evaporation (mechanism 3) will depend on the level of moisture saturation in the air and the ventilation rate of the target area. In other words, cities located in humid climates will benefit less from these techniques than those in dry ones. Similarly, poorly ventilated and confined urban areas are likely to saturate the vapor pressure deficit and will therefore also benefit less from evaporative cooling than more open and highly ventilated areas such as rooftops.

Accurate knowledge of this parameter requires a numerical simulation and is a priori out of reach for simple GIS tools.
- **Building typology (mechanism 2)**

Even though energy efficiency measures can be applied blindly, information on the building stock and its energy performance is useful to target the least efficient buildings first.

More and more cities already have such databases at the building level. However, more detailed knowledge may be required to better model anthropogenic heat release to the atmosphere at the pedestrian or street level (e.g. position and type of air conditioners).

## Conclusion

The physical mechanisms on which cooling techniques rely were analyzed on the basis of the energy balance of an urban surface and volume. This made it possible to classify the techniques according to these mechanism(s).

On this basis, the most influential parameters for each technique were identified and their correspondence with geographical data established where available. Three parameters appear to be of primary importance and affect a large number of cooling techniques: solar irradiance, existing materials and underground infrastructure. Solar irradiance is the most important of these, especially because it is the strongest inbound flux during the day. Knowledge of existing materials and underground infrastructure is also very useful, although both provide information relevant more locally. Other parameters, relevant for a narrower spectrum of cooling techniques, have also been identified.

Among the data identified, it appears in particular that access to a DEM of the city studied is very important, since it is key to gaining access to solar irradiance and sky view factor in particular.

## Acknowledgments

The authors acknowledge the Parisian Urban Planning Agency (APUR) and Paris City Hall for the solar irradiance data and infrared photograph. This research was funded by the Urban Materials for Cool Cities (UMat4CC) project (ANR-18-CE22-015-01).